\documentstyle[aps,floats,epsf]{revtex}
\def\heimo{Hei\-del\-berg--Mos\-cow ex\-pe\-ri\-ment}
\def\onbb{0$\nu\beta\beta$}

\begin{document}
\renewcommand{\topfraction}{1.0}

\draft
\title{Limits on the Majorana neutrino mass in the 0.1 eV range}
\author{L. Baudis, A. Dietz, G. Heusser, H.V. Klapdor--Kleingrothaus, I.V. Krivosheina 
\thanks{On leave from the Radiophysical Research Institute
  (NIRFI), Nishnij Novgorod, Russia},
St. Kolb, B. Majorovits, V.F. Melnikov $^*$, H. P\"as, F. Schwamm and H. Strecker}
\address{Max--Planck--Institut f\"ur Kernphysik, \\
P.O.Box 10 39 80, D--69029 Heidelberg, Germany}
\author{V. Alexeev, A. Balysh, A. Bakalyarov, S.T. Belyaev, V.I. Lebedev and S. Zhukov}
\address{Russian Science Centre Kurchatov Institute, \\
123 182 Moscow, Russia}
\maketitle
\begin{abstract}
The \heimo{} gives the most stringent limit on the Majorana neutrino
mass. After 24 kg yr of data with pulse shape measurements, we set a
lower limit on the half-life of the \onbb{}-decay in $^{76}$Ge of  
${\rm T}_{1/2}^{0\nu} \geq 5.7 \times 10^{25} {\rm~ yr}$ at 90\% C.L., 
thus excluding an effective Majorana neutrino mass greater than 0.2 eV. 
This allows to set strong constraints on degenerate neutrino mass models.
\end{abstract}


Neutrinoless double beta (\onbb{}) decay is an extremely sensitive
tool to probe  theories beyond the standard model (see \cite{klap}). 
While the standard
model exactly conserves B-L, \onbb{}-decay violates lepton number,
and B-L, by two 
units. The simplest mechanism which can induce \onbb{}-decay is the
exchange of a Majorana neutrino between the decaying
neutrons. Alternatively, any theory that contains lepton number
violating interactions can lead to the process. Independently of the
underlying mechanism, an observation of the \onbb{}-decay would be an
evidence for a nonzero Majorana neutrino mass \cite{schechter}.
There are several indications for nonzero neutrino masses, the most
stringent ones come from solar and atmospheric neutrino experiments. 
In particular, the confirmation by Super Kamiokande of the atmospheric 
neutrino deficit \cite{superK}, provides strong
evidence for neutrino oscillations, although also other solutions are
possible \cite{nunokawa}. If a neutrino as a hot dark matter
(HDM) component is taken into account, then fitting the atmospheric,
solar and HDM scales with three neutrinos is only possible in the
degenerate mass scenario, where all neutrinos have nearly the same
mass, in the order of ${\cal O}$(eV) \cite{caldwell}. 
This would
lead to an amplitude for \onbb{}-decay mediated by the
neutrino mass which is
accessible by the present sensitivity of the \heimo{}.

The \heimo{} operates five p--type HPGe detectors in the Gran Sasso Underground Laboratory.
The Ge crystals were grown out of
19.2 kg of 86\% enriched $^{76}$Ge material. 
The total active mass of the detectors is 10.96 kg,
corresponding to 125.5 mol of $^{76}$Ge, the presently largest source
strength of all double beta experiments.
Four detectors are placed in a common 30 cm thick lead shielding in a
radon free nitrogen atmosphere, surrounded by 10 cm of boron-loaded
polyethylene and with two layers of 1 cm thick scintillators on top.
The remaining detector is situated in a separate box with 27 cm electrolytical
copper and 20 cm lead shielding, flushed with gaseous nitrogen and
with 10
cm of boron-loaded polyethylene below the box. A detailed description
of the experiment and its background is given in \cite{heimo1}.
For a further reduction of the already very low background of the
experiment, a pulse shape analysis (PSA) method was developed
\cite{heimo2}. The analysis distinguishes between multiple
scattered interaction in the Ge crystal, so called multiple site
events (MSE) and pointlike interactions, i.e. single site events (SSE).
Since double beta decay events belong to the SSE category, the method
allows to effectively reduce the background of multiple Compton
scattered photons. The probability of correct detection for a SSE is
75\%, and 74\% for a MSE \cite{heimo2}.

\begin{figure}
\epsfxsize=15cm
\centerline{\epsffile{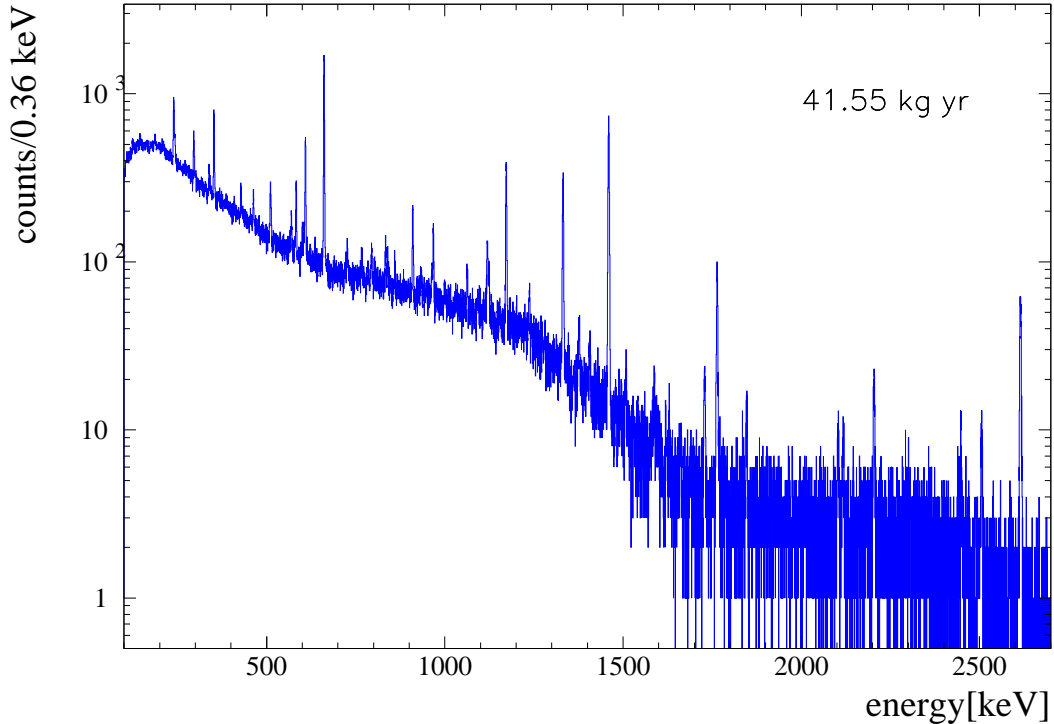}}
\caption{Sum spectrum of all detectors of the Heidelberg-Moscow
  experiment  with 41.55 kg yr for the energy region between 100 and 2700 keV.}
\label{sumspec}
\end{figure}

Figure \ref{sumspec} shows the total spectrum of
the five enriched detectors of the \heimo{}, with a statistical
significance of 41.55 kg yr. Due to the good energy resolution of the
detectors, the $\gamma$-activities can be easily
identified via their specific lines. The neutron background can be
estimated from measurements with and without the neutron shielding,
while the effect of muons can be deduced from the scintillator
measurements in coincidence with Ge detectors. In the region of
interest for the \onbb{}-decay in $^{76}$Ge (Q-value = 2038.56
$\pm$0.32 keV \cite{hyka91})
the dominant background originates from the Compton continuum of the
$^{208}$Tl line (2614 keV), the summed $^{60}$Co line (2505 keV) and
some $^{214}$Bi lines (2118.5 keV, 2204 keV, 2246 keV) (all together
about 60\% of the total background),
from neutron-induced (about 30\%) and muon-induced (about 10\%)
events.

For the evaluation of the \onbb{}-decay we consider both data sets,
with and without pulse shape analysis. We see in none of them an
indication for a peak at the Q-value of the \onbb{}-decay.
The total spectrum of the five detectors with a statistical
significance of 41.55 kg yr contains all the data with exception of
the first 200 d of measurement of each detector.
The interpolated energy resolution at the energy of the hypothetical
\onbb{}-peak is (3.85$\pm$0.16) keV. To estimate the expected
background in the \onbb{} region, we take the energy interval from
2000 to 2080 keV. The number of expected events in the peak region is
(78$\pm$3) events, the number of measured events in the 3$\sigma$ peak 
interval centered at 2038.56 keV is 68. This number is below the
expectation and might be due to peaks in the background interval,
which are however too weak to be identified.
To extract a half-life limit for the \onbb{}-decay we make the
assumption, as recommended by \cite{pdg96}, that the number of
measured events equals the background expectation.
With the achieved energy resolution, the number of excluded events in
the 3$\sigma$ peak region is 15.9 (9.52) with 90\% C.L. (68\% C.L.),
resulting in a half-life limit of (for the 0$^+ \rightarrow$ 0$^+$
transition):

\begin{eqnarray*}
{\rm T}_{1/2}^{0\nu} \geq 1.3 \times 10^{25} {\rm~ yr} \;\;\; 90\% {\rm~ C.L.}\\
{\rm T}_{1/2}^{0\nu} \geq 2.1 \times 10^{25} {\rm~ yr} \;\;\; 68\% {\rm~ C.L.}
\end{eqnarray*}

We consider now the data with pulse shape measurements, with a
statistical significance of 24.16 kg yr and an energy resolution at
2038.56 keV of (4.2$\pm$0.17) keV. The expected number of events 
from the background left and right of the peak is (13$\pm$1)
events, the measured number of events in the 3$\sigma$ peak region is
7. Considering again the number of expected events instead of the number of 
measured ones \cite{pdg96}, we can exclude 7.17 (4.17) events with 90\%
C.L. (68 \% C.L.). The limit on the half-life is:

\begin{eqnarray*}
{\rm T}_{1/2}^{0\nu} \geq 1.6 \times 10^{25} {\rm~ yr} \;\;\; 90\% {\rm~ C.L.}\\
{\rm T}_{1/2}^{0\nu} \geq 2.8 \times 10^{25} {\rm~ yr} \;\;\; 68\% {\rm~ C.L.}
\end{eqnarray*}

Obviously the pulse shape data are now not only competitive with the
complete data set, but they deliver more stringent lower limits on the 
half-life of the \onbb{}-decay. The pulse shape analysis reduces the
background in the interesting energy region by a factor of 3
[background index without PSA: (0.18$\pm$0.02) events/(kg yr keV),
with PSA: (0.06$\pm$0.02) events/(kg yr keV)].
This reduction factor is due to the large fraction of multiple Compton 
scattered events in the \onbb{}-decay region.

\begin{figure}[t]
\epsfxsize=15cm
\centerline{\epsffile{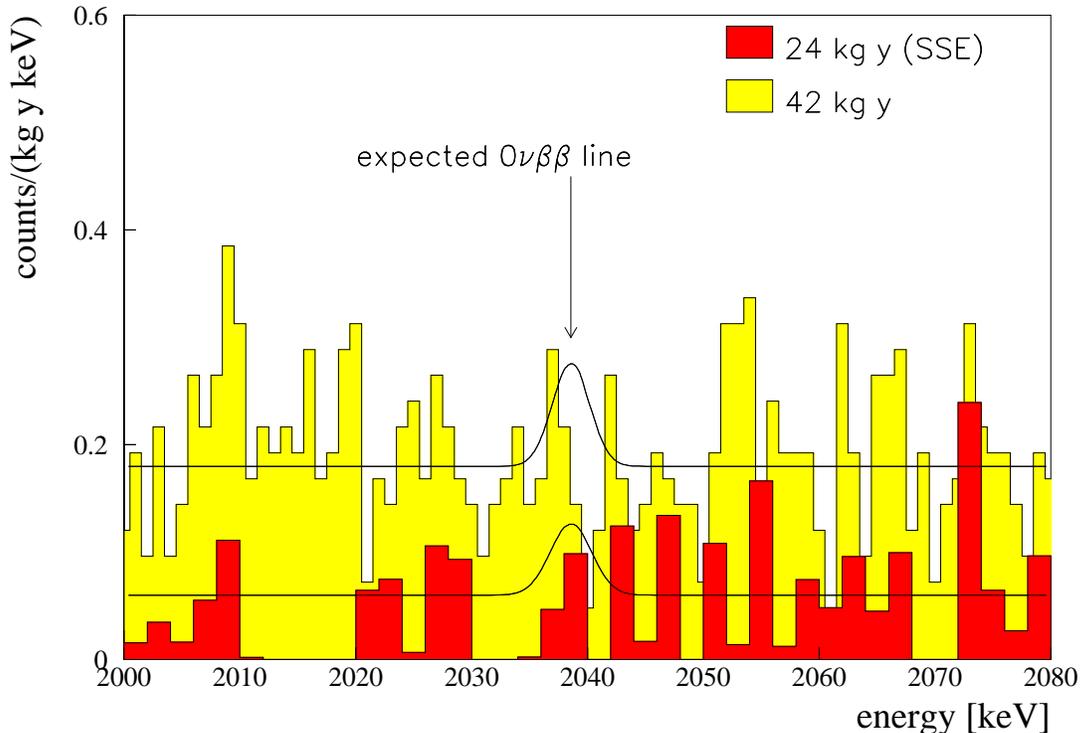}}
\caption{Sum spectrum of all five detectors with 41.55 kg yr and SSE
  spectrum with 24.16 kg yr in the region of interest for the
  \onbb{}-decay. The curves correspond to the excluded signals with
  ${\rm T}_{1/2}^{0\nu} \geq 1.3 \times 10^{25} {\rm~ yr}$ (90\%
  C.L.) and ${\rm T}_{1/2}^{0\nu} \geq 1.6 \times 10^{25} {\rm~ yr}$ (90\% C.L.) }
\label{both}
\end{figure}

Figure \ref{both} shows the combined spectrum of the five detectors after 41.55
kg yr and the SSE spectrum, corrected for the detection efficiency,
after 24.16 kg yr. The solid lines represent the exclusion limits for
the two spectra at the 90\% C.L.

In addition we report the limits obtained by evaluating the SSE
data with the new method proposed by the Particle Data Group 98 \cite{pdg98}.
The number of excluded events for an observation of 7 events and a
background expectation of 13 is 2.07 (0.47) with 90\% C.L. (68\% C.L.)
(see Table III and V in \cite{feldm98}). The resulting upper limits
for the half-life of the \onbb{}-decay are the following:

\begin{eqnarray*}
{\rm T}_{1/2}^{0\nu} \geq 5.7 \times 10^{25} {\rm~ yr} \;\;\; 90\% {\rm~ C.L.}\\
{\rm T}_{1/2}^{0\nu} \geq 2.5 \times 10^{26} {\rm~ yr} \;\;\; 68\% {\rm~ C.L.}
\end{eqnarray*}

The sensitivity of the experiment, as defined in \cite{feldm98},
is again obtained by setting the measured number of events equal to the
expected background. With 7.51 (4.71) excluded events at 90\% C.L
(68\% C.L.), the half-life limit is:

\begin{eqnarray*}
{\rm T}_{1/2}^{0\nu} \geq 1.6 \times 10^{25} {\rm~ yr} \;\;\; 90\% {\rm~ C.L.}\\
{\rm T}_{1/2}^{0\nu} \geq 2.5 \times 10^{25} {\rm~ yr} \;\;\; 68\% {\rm~ C.L.}
\end{eqnarray*}
  
\begin{figure}
\epsfxsize=15cm
\centerline{\epsffile{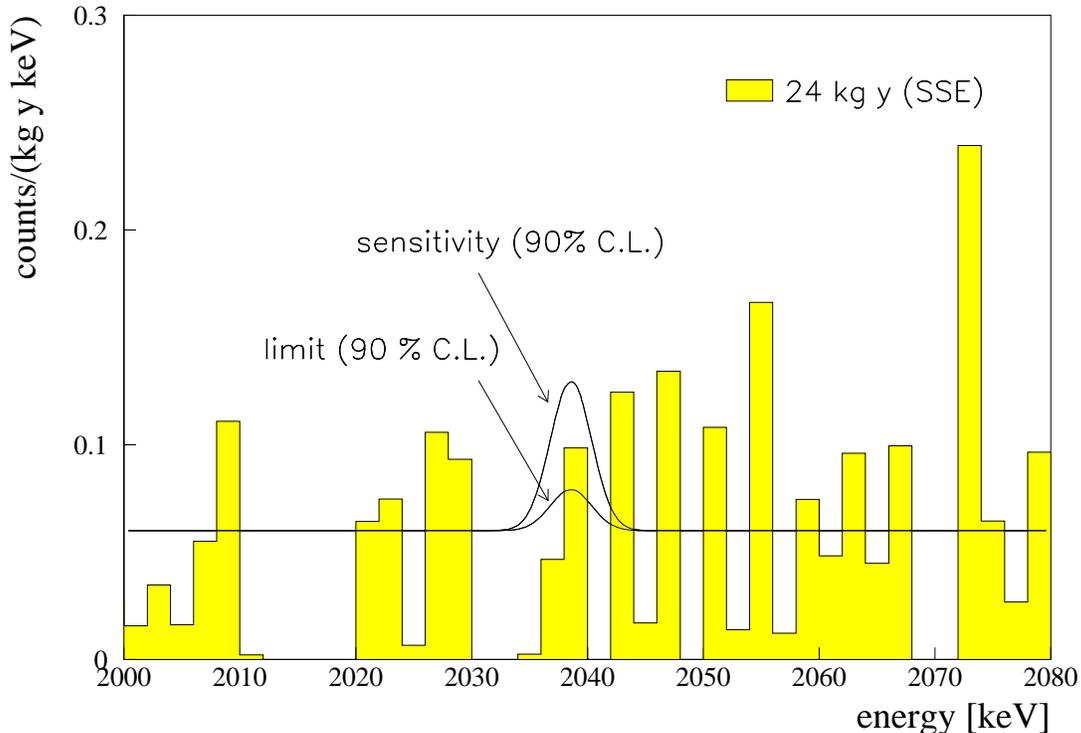}}
\caption{SSE spectrum after 24.16 kg yr and the corresponding excluded
  signal and sensitivity of the experiment, at 90\% C.L. after \protect{\cite{pdg98}}. }
\label{sse}
\end{figure}

Figure \ref{sse} shows the SSE spectrum and the excluded peaks 
(upper limit and sensitivity) with 90\% C.L.

Using the matrix elements of \cite{staudt} and neglecting right-handed 
currents, we can
convert the lower half-life limit into an upper limit on the effective 
Majorana neutrino mass. The obtained limits are shown in Table \ref{n_mass}.
The \heimo{} is thus giving the up to now most stringent upper limit on the
Majorana neutrino mass, of 0.2 eV at 90\% C.L. (0.1 eV at 68\% C.L.) 
\cite{pdg98}. The sensitivity of the experiment is 0.38 eV at 90\% 
C.L. \cite{pdg98}.

Table \ref{bb_exp} gives a list of the limits from the presently most sensitive
experiments on double beta decay and a comparison between 
different matrix element calculations. It should be mentioned, that
the calculations of \cite{engel} do not use a realistic
nucleon-nucleon force, those of \cite{caurier} use a too small
configuration space (leaving out the important spin-orbit partners) and the recent
calculation of \cite{simko} does not fulfill the Ikeda sum rule. 
This favours the matrix elements of \cite{staudt,tomoda} (for a
discussion see also \cite{klap}).

The limit on the effective Majorana neutrino mass given by the
\heimo{} is about an order of magnitude lower than for other
double beta experiments. This is the result of the high source strength, 
good energy resolution, high material purities combined with the
efficiency of the pulse shape analysis and last, but not
least, of the excellent long-term stability of the experiment.

After about another five years of measurement, assuming no other
improvement in the present background index, the 
\heimo{} will be able to explore the half-life of the \onbb{}-decay  
up to some 10$^{26}$ yr in the best case.
For a significant improvement of the experimental
sensitivity, a much lower background counting rate and a higher source
strength are required. 

A major step forward in this field would bring
the new project proposed by our group, GENIUS (GErmanium in liquid
NItrogen Underground Setup) \cite{klap,hellmig}, which would operate enriched Ge crystals
directly in liquid nitrogen. The goal is to reduce the background by
about
three orders of magnitude by removing essentially all materials from
the vicinity of the measurement crystals. 
It was shown that Ge detectors work reliably in liquid nitrogen
\cite{hellmig,baudis}, also detailed Monte Carlo simulations of the
various expected background components \cite{baudis} confirm the
possibility to obtain a counting rate of 0.3 events/(t yr keV) in the
\onbb{}-region. GENIUS is conceived to test the neutrino mass down to
the level of 0.01 eV and lower.
 
In conclusion, already in the present stage, the \heimo{} is setting the 
most stringent limit on the Majorana neutrino mass, allowing to test
the predictions of degenerate neutrino mass models. 
For example, in models which try to accommodate the solar- and atmospheric neutrino
problems while considering the neutrino as a hot dark matter candidate 
with a mass of a few eV, the small angle MSW solution is practically ruled out
\cite{minakata96}. For the large angle MSW solution, an effective
Majorana neutrino mass smaller than 0.2 eV would need an unnatural
fine tuning to account for a relevant neutrino mass in a 
mixed hot and cold dark matter cosmology \cite{minakata98}. 

This, as well as other implications
for physics beyond the standard model, like left-right symmetric 
models, supersymmetry, leptoquarks and compositeness will be discussed 
in detail elsewhere.

\begin{table}
\caption{Limits on the effective Majorana neutrino mass from the
  \onbb{}-decay of $^{76}$Ge for the matrix elements from \protect{\cite{staudt}}.}
\label{n_mass}
\begin{tabular}{lccc}
& T$_{1/2}^{0\nu}$ [yr] & $\langle m \rangle$ [eV] & C.L. [\%]\\\hline
Full data set & $\geq 1.3 \times 10^{25}$ & $\leq$ 0.43 & 90\\
 & $\geq 2.1 \times 10^{25}$ & $\leq$ 0.33 & 68\\
SSE data after \cite{pdg96} & $\geq 1.6 \times 10^{25}$ & $\leq$ 0.38 & 90\\
&$\geq 2.8 \times 10^{25}$ & $\leq$ 0.29 & 68\\
SSE data after \cite{pdg98} &$\geq 5.7 \times 10^{25}$ & $\leq$ 0.20 & 90\\
&$\geq 2.5 \times 10^{26}$ & $\leq$ 0.10 & 68\\
Sensitivity & $\geq 1.6 \times 10^{25}$ & $\leq$ 0.38 & 90\\
&$\geq 2.5 \times 10^{25}$ & $\leq$ 0.30 & 68\\
\end{tabular} 
\end{table}

\begin{table}
\caption{Limits on the half-lifes of the \onbb{}-decay and on the
  effective Majorana neutrino mass for different matrix element
  calculations for the best existing experiments with half-life limits $>$
  10$^{21}$ yr.}
\label{bb_exp}
\begin{tabular}{llll}
Isotope & T$_{1/2}^{0\nu}$ [yr] & C.L. [\%] & $\langle m \rangle$ [eV] \\\hline
$^{48}$Ca &  $\geq$ 9.5$\times$10$^{21}$  & 76 \cite{Ke91}    & $\leq$ 13 \cite{staudt}\\
$^{76}$Ge limit &  $\geq$ 5.7$\times$10$^{25}$  & 90   & $\leq$ 0.20
\cite{staudt}, 0.56 \cite{caurier}, 0.52 \cite{engel}, 0.19
\cite{tomoda}, 0.4 eV \cite{simko}\\
$^{76}$Ge sensitivity &$\geq$ 1.6$\times$10$^{25}$  & 90   & $\leq$
0.38 \cite{staudt}, 1.07 \cite{caurier}, 0.93 \cite{engel}, 0.36
\cite{tomoda}, 0.77 \cite{simko}\\
$^{82}$Se &  $\geq$ 9.5$\times$10$^{21}$  & 90 \cite{arnold98}& $\leq$
7.9 \cite{staudt}, 15.7 \cite{caurier}, 24 \cite{engel}, 8 \cite{tomoda}\\
&  $\geq$ 2.7$\times$10$^{22}$  & 68 \cite{elliott92}& $\leq$ 4.7
          \cite{staudt}, 9.4 \cite{caurier}, 14.4 \cite{engel}, 4.75 \cite{tomoda}\\ 
$^{100}$Mo & $\geq$ 5.2$\times$10$^{22}$  & 68 \cite{kudomi98} &
$\leq$ 4.9 \cite{staudt}, 4.34 \cite{engel}, 2.2 \cite{tomoda}, 2.2 \cite{simko}\\
$^{116}$Cd & $\geq$ 3.2$\times$10$^{22}$  & 90 \cite{danevich99}& $\leq$ 3.9 \cite{staudt}\\
$^{130}$Te & $\geq$ 5.6$\times$10$^{22}$  & 90 \cite{allessan98}&
$\leq$ 2.9 \cite{staudt}, 3.43 \cite{engel}, 3.1 \cite{tomoda}, 2.88 \cite{simko}\\ 
$^{136}$Xe & $\geq$ 4.4$\times$10$^{23}$  & 90 \cite{luescher98}&
$\leq$ 2.2 \cite{staudt}, 5.2 \cite{caurier}, 2.7 \cite{engel}, 1.8 \cite{tomoda}\\ 
$^{150}$Nd & $\geq$ 1.22$\times$10$^{21}$ & 90 \cite{desilva97}& $\leq$ 5.2 \cite{staudt}\\
\end{tabular} 
\end{table}

\acknowledgments 
The Heidelberg--Moscow experiment was supported by the 
Bundesministerium f\"ur Forschung und Technologie der Bundesrepublik 
Deutschland, the State Committee of Atomic Energy of Russia and the 
Istituto Nazionale di Fisica Nucleare of Italy. L.B. was supported by 
the Graduiertenkolleg of the University of Heidelberg.

\end{document}